\documentclass[10 pt, letter]{article}
\usepackage{graphicx}
\usepackage{subfigure}
\usepackage{caption}
\usepackage{amsmath}
\usepackage{amsthm}
\usepackage{enumerate}
\usepackage{subfig}
\usepackage{float}
\usepackage{amsfonts}
\usepackage[spanish]{babel}
\usepackage[latin1]{inputenc}
\usepackage{amssymb}
\usepackage{anysize}
\usepackage{helvet}
\usepackage{setspace}
\usepackage{multirow}
\usepackage{tabularx}
\usepackage{tabulary}
\usepackage{longtable}
\usepackage{float}
\usepackage{mathtools}
\usepackage[none]{hyphenat}

\usepackage{cite}
\title{\bf Quantization of the 1-D forced harmonic oscillator in the space ($x, v$) }
\author{ Gustavo V. L\'opez\footnote{gulopez@cencar.udg.mx}, and O.J.P. Bravo\footnote{hase12@hotmail.com}.
	\\
	\\
 Departamento de F\'{i}sica, Universidad de Guadalajara,\\
 Blvd. Marcelino Garc\'{i}a Barragan y Calzada Ol\'{i}mpica, \\ CP 44200, Guadalajara, Jalisco, M\'exico, \\ 
 \\  
 }
\begin{document}
\maketitle

\begin{abstract}
\noindent
The quantization of the forced harmonic oscillator is studied with the quantum variable ($x,\hat v$), with the commutation relation $[x,\hat v]=i\hbar/m$,
 and using a Shr\"odinger's like equation on these variable, and associating a linear operator to a constant of motion $K(x,v,t)$ 
of the classical system,  The comparison with the  quantization
in the space ($x,p$) is done with the usual Schr\"odinger's equation for the Hamiltonian $H(x,p,t)$, and with the commutation relation 
$[x,\hat p]=i\hbar$. It is found that for the non resonant case, both forms of quantization brings about the same result. However, for
the resonant case, both forms of quantization are different, and the probability for the system to be in the exited state for the ($x,\hat v$) 
quantization has less oscillations than the ($x,\hat p$) quantization, the average energy of the system is higher in ($x,\hat p$) quantization 
than on the $(x,\hat v$) quantization, and the Boltzmann-Shannon entropy on the ($x,\hat p$) quantization is higher than on the ($x,\hat v$) quantization.  
\end{abstract}
\vskip1cm
\newpage
\section{Introduction}
The usual quantum mechanics formulation is done in the space ($x,\hat p$)  \cite{neumann}, where $\hat p=-i\hbar\partial/\partial x$ is the linear operator associated to the 
classical generalized linear momentum of the motion of a particle of mass ``m," where the commutation relation $[x,\hat p]=i\hbar$ \cite{born} is satisfied. A linear operator is
associated to the classical Hamiltonian, $\widehat{H}(x,\hat p,t)$, to form the so called Schr\"odinger's equation \cite{schr}
\begin{equation}\label{sh}
i\hbar\frac{\partial\Psi}{\partial t}=\widehat{H}(x,\hat p,t)\Psi,
\end{equation}  
where $\Psi=\Psi(x,t)$ is the wave function. This formulation has have enormous success to explain and to predict many microscopic behavior of the nature \cite{spielman}. 
However, despite this enormous success, Hamiltonian-Lagrangian mathematical formulation has some details, even for 1-D problem where one knows that the 
Lagrangian (therefore the Hamiltonian) always exists \cite{darboux}. First, from the expression to obtain the generalized linear momentum given the Lagragian, $L(x,\dot x,t)$ 
for the system,
\begin{equation}
p(x,\dot x,t)=\frac{\partial L}{\partial\dot x},
\end{equation} 
it is not always possible to obtain explicitly $\dot x=\dot x(x,p,t)$ to be able to get the explicit expression for the Hamiltonian from the Legrandre's transformation \cite{goldstein},
\begin{equation}
H(x,p,t)={\dot x}(x,p,t)p-L(x,{\dot x}(x,p,t),t).
\end{equation}
Second, when one is dealing with classical dissipative systems \cite{lpez2007},
\begin{equation}
\frac{d(m\dot x)}{dt}=F(x,\dot x),
\end{equation} 
either it is not possible to find its Hamiltonian, or two different Hamiltonians are possible to find for the system \cite{lll,dodman1,dodman2,lop3,mont}. Last one, for those problems of variable mass systems,
\begin{equation}
\frac{d\bigl(m(x,\dot x,t)\dot x\bigr)}{dt}=F(x),
\end{equation}
which are not invariant under Galileo's transformations and Sommerfeld modification is not consistent, to find the Hamiltonian for this system \cite{lopez} requires to start
from the ``Inverse Problem of the Mechanics."
\vskip1pc\noindent
Therefore, one has the necessity to find some extension of the known quantization arised from the\break\hfil Hamilton-Lagrangian approach. In this way, there is already a 
proposition \cite{lok,lok2}  of using a function $K(x,v,t)$ that could be a constant of motion of the classical system, and to associate a linear operator to the velocity of the form
\begin{equation}
\hat v=-i\frac{\hbar}{m}\frac{\partial}{\partial x},
\end{equation}  
such that $[x,\hat v]=i\hbar/m$, and to associate a linear operator
\begin{equation}
K(x,v,t)\quad\longrightarrow\quad\widehat{K}(x,\hat v,t),
\end{equation}
which can be used to form the Shr\"odinger's like equation 
\begin{equation}\label{sk}
i\hbar\frac{\partial\Psi}{\partial t}=\widehat{K}(x,\hat v,t)\Psi.
\end{equation}
In this paper, approaches (\ref{sh}) and (\ref{sk}) are used to study the 1-D forced harmonic oscillator to determine whether or not there is a difference on the
quantization, and hopefully to see if the approach (\ref{sk}) could have with these result and experimental verification.
\section{Analytical Approach for $K(x,v,t)$}
The forced harmonic oscillator is classically characterized by Newton's equation
\begin{equation}
\frac{d(m\dot x)}{dt}=-m\omega_0^2x+\alpha\cos(\omega t+\varphi),
\end{equation} 
where ``m'' is the mass of the particle, $\omega_0$ is the natural frequency of oscillation (when $\alpha=0$), and $\alpha$ is the amplitude of the 
forced force. The well known solution of this problem is
\begin{equation}\label{xsol}
x(t)=\begin{cases}C_1\cos\omega_0t+C_2\sin\omega_0t+\frac{\alpha\cos(\omega t+\varphi)}{m(\omega_0^2-\omega^2)}\quad \omega\not=\omega_0\\ \\
C_1\cos\omega_0t+C_2\sin\omega_0t+\frac{\alpha\sin(\omega_0 t+\varphi)}{2m\omega_0}t\quad\omega=\omega_0\end{cases},
\end{equation}
where one has the non resonant case ($\omega\not=\omega_0$) and the resonant case ($\omega=\omega_0$). The velocity is known by making the differentiation of (\ref{xsol}) 
with respect the time, and the constants $C_1$ and $C_2$ are determined by the initial condition ($x(0), v(0)$). For the non resonant case, these constants are
\begin{subequations}
\begin{eqnarray}
C_1&=&x\cos\omega_0t-\frac{v}{\omega_0}\sin\omega_0t\nonumber\\
& &-\frac{\alpha}{m(\omega_0^2-\omega^2)}\left\{\cos(\omega t+\varphi)\cos\omega_0t+\frac{\omega}{\omega_0}\sin(\omega t+\varphi)\sin\omega_0t\right\}
\end{eqnarray}
and
\begin{eqnarray}
C_2&=&x\sin\omega_0t+\frac{v}{\omega_0}\cos\omega_0t\nonumber\\
& &-\frac{\alpha}{m(\omega_0^2-\omega^2)}\left\{\cos(\omega t+\varphi)\sin\omega_0t-\frac{\omega}{\omega_0}\sin(\omega t+\varphi)\cos\omega_0t\right\}
\end{eqnarray}
\end{subequations}
For the resonant case ($\omega=\omega_0$), one has
\begin{subequations}
\begin{eqnarray}
C_1&=&x\cos\omega_0t-\frac{v}{\omega_0}\sin\omega_0t\nonumber\\
& &+\frac{\alpha}{2m\omega_0}\left\{-t\sin(\omega_0 t+\varphi)\cos\omega_0t+t\cos(\omega_0t+\varphi)\sin\omega_0t+\frac{1}{\omega_0}\sin(\omega_0 t+\varphi)\sin\omega_0t\right\}
\end{eqnarray}
and
\begin{eqnarray}
C_2&=&x\sin\omega_0t+\frac{v}{\omega_0}\cos\omega_0t\nonumber\\
& &-\frac{\alpha}{2m\omega_0}\left\{t\sin(\omega_0 t+\varphi)\sin\omega_0t+t\cos(\omega_0t+\varphi)\cos\omega_0t+\frac{1}{\omega_0}\sin(\omega_0 t+\varphi)\cos\omega_0t\right\}
\end{eqnarray}
\end{subequations}
Now, by choosing a constant of motion of the form
\begin{equation}
K_{\alpha}^{(nr,r)}(x,v,t)=\frac{1}{2}m\omega_0^2\biggl(C_1^2+C_2^2\biggr),
\end{equation}
where `nr" means non resonant and ``r" means resonant,  it follows that
\begin{equation}
\lim_{\alpha\to 0}K_{\alpha}^{(nr,r)}(x,v,t)=\frac{1}{2} mv^2+\frac{1}{2}m\omega_0^2x^2,
\end{equation} 
which represents the usual energy of the harmonic oscillator, independently of the non resonant case or resonant case. This constant of motion can be written as
\begin{equation}
K_{\alpha}^{(nr,r)}(x,v,t)=K_0(x,v)+W_{\alpha}^{(nr,r)}(x,v,t),
\end{equation}
where $K_0$ and $W_{\alpha}^{(nr,r)}$ are defined as
\begin{equation}
K_0(x,v)=\frac{1}{2} mv^2+\frac{1}{2}m\omega_0^2x^2,
\end{equation}
\begin{equation}
W_{\alpha}^{(nr)}(x,v,t)=\frac{1}{2}m\omega_0^2\left[A^2\cos^2(\omega t+\varphi)-2Ax\cos(\omega t+\varphi)+B^2\sin^2(\omega t+\varphi)+\frac{2Bv}{\omega_0}\sin(\omega t+\varphi)\right],
\end{equation}
and
\begin{eqnarray}\label{wr}
W_{\alpha}^{(r)}(x,v,t)&=&\frac{1}{2}m\omega_0^2\biggl[a^2(t)-\frac{2a(t) v}{\omega_0}\cos(\omega_0t+\varphi)-\frac{2b v}{\omega_0}\sin(\omega_0t+\varphi)\nonumber\\
& &+2a(t) b\cos(\omega_0t+\varphi)\sin(\omega_0t+\varphi)-2a(t) x\sin(\omega_0t+\varphi)+b^2\sin^2(\omega_0t+\varphi)\biggr],
\end{eqnarray}
where one has made the definitions
\begin{subequations}
\begin{equation}
A=\frac{\alpha}{m(\omega_0^2-\omega^2)},\quad\quad B=\frac{\alpha\omega}{m\omega_0(\omega_0^2-\omega^2)}
\end{equation}
and
\begin{equation}\label{at}
a(t)=\frac{\alpha t}{2m\omega_0},\quad\quad b=\frac{\alpha}{2m\omega_0^2}.
\end{equation}
\end{subequations}
To solve equation (\ref{sk}), one observes that the eigenvalues problem for the operator $\widehat{K}_0$,
\begin{equation}
\widehat{K}_0(x,\hat v)\Phi=E\Phi,
\end{equation}
has exactly the same solution of the that one given by the Hamiltonian problem, $(\hat{p}^2/2m+m\omega_0^2 x^2/m)\Phi=E\Phi$ where
the solution is the set $\{E_m^{(0)}, \Phi_n(x)\}_{n\ge 0}$,
\begin{subequations}\label{eigenho}
\begin{equation}
E_n^{(0)}=\hbar\omega_0(n+1/2)
\end{equation}
and
\begin{equation}
\Phi_n(x)=A_ne^{-\xi^2/2}H_n(x),\quad\quad \xi=\sqrt{\frac{m\omega_0}{\hbar}}~x,\quad A_n=\left(\frac{m\omega_0}{\pi\hbar}\right)^{1/4}\frac{1}{\sqrt{2^n n!}}.
\end{equation}
\end{subequations}
Using Dirac's notation \cite{dirac}, where $\Phi_n(x)=\langle x|n\rangle$, with $|n\rangle$ characterizing the nth-state, and then one has the eigenvalue problem written as
\begin{equation}\label{ko}
\widehat{K}_0|n\rangle=E_n^{(0)}|n\rangle.
\end{equation}
Therefore, one can propose the solution of the Shr\"odinger's equation(\ref{sk}) with the operator constant of motion $\widehat{K}$,
\begin{equation}
i\hbar\frac{\partial |\Psi(t)\rangle}{\partial t}=\left\{\widehat{K}_0(x,\hat v)+W_{\alpha}^{(nr,r)}(x,\hat v,t)\right\}|\Psi(t)\rangle,
\end{equation}
 of the form
 \begin{equation}
|\Psi(t)\rangle=\sum_{n=0}^{\infty}C_n(t)|n\rangle.
\end{equation}
Taking into consideration (\ref{ko}), the orthogonality of the states ($\langle m|n\rangle=\delta_{mn}$), one obtains the following equation for the coefficients
\begin{equation}\label{cq}
i\hbar\dot C_m(t)=E_m^{(0)}C_m(t)+\sum_{n=0}^{\infty}C_n(t) W_{mn}^{(nr,r)}(t),
\end{equation}
where $W_{mn}^{(nr,r)}(t)$ represents the matrix element
\begin{equation}
W_{mn}^{(nr,r)}(t)=\langle m|W_{\alpha}^{(nr,r)}|n\rangle.
\end{equation}
The equation (\ref{cq}) can be simplified using the new variable
\begin{equation}
C_k(t)=e^{-iE_k^{(0)} t/\hbar}D_k(t).
\end{equation}
The equations for these new coefficients are
\begin{equation}\label{do}
i\hbar {\dot D}_m(t)=\sum_{n=0}^{\infty} e^{i\omega_{mn}t}D_n(t) W_{mn}^{(nr,r)}(t),
\end{equation}
where $C_k(0)=D_k(0)$ and the probability to find the system in the state $|k\rangle$ is $|C_k(t)|^2=|D_k(t)|^2$. Matrix elements are much easier to 
calculate by using the non Hermitian ascent  `$a^{\dagger}$'' and descent ``$a$'' operators,
\begin{equation}
a=\sqrt{\frac{m\omega_0}{2\hbar}}~x+i\sqrt{\frac{m}{2\omega_0\hbar}}~\hat v,\quad\quad a^{\dagger}=\sqrt{\frac{m\omega_0}{2\hbar}}~x-i\sqrt{\frac{m}{2\omega_0\hbar}}~\hat v,
\end{equation}
with the knows properties \cite{burk}
\begin{subequations}
\begin{equation}
[a,a]=[a^{\dagger},a^{\dagger}]=0,\quad\quad [a,a^{\dagger}]=1,
\end{equation}
and
\begin{equation}
a^{\dagger}|n\rangle=\sqrt{n+1}|n+1\rangle,\quad\quad a|n\rangle=\sqrt{n}|n-1\rangle.
\end{equation}
\end{subequations}
For non resonant case (nr), after calculating the matrix elements , using the orthogonality of the states, and making some rearrangements, one gets the equations for the real and imaginary
parts of the coefficients, $D_k(t)=X_k(t)+iY_k(t)$, as
\begin{subequations}\label{eqnr}
\begin{eqnarray}
{\dot X}_k&=& -c\sqrt{k}\bigl[\cos(\omega t+\varphi)\cos\omega_0t~Y_{k-1}+\cos(\omega t+\varphi)\sin\omega_0t~X_{k-1}\bigr]\nonumber\\
& &-c\sqrt{k+1}\bigl[\cos(\omega t+\varphi)\cos\omega_0t~Y_{k+1}-\cos(\omega t+\varphi)\sin\omega_0t~X_{k+1}\bigr]\nonumber\\
& &+d\sqrt{k}\bigl[\sin(\omega t+\varphi)\cos\omega_0t~X_{k-1}-\sin(\omega t+\varphi)\sin\omega_0t~Y_{k-1}\bigr]\nonumber\\
& &-d\sqrt{k+1}\bigl[\sin(\omega t+\varphi)\cos\omega_0t~X_{k+1}+\sin(\omega t+\varphi)\sin\omega_0t~Y_{k+1}\bigr]\nonumber\\
& &+a_1\cos^2(\omega t+\varphi)~Y_k+b_1\sin^2(\omega t+\varphi)~Y_k
\end{eqnarray}
\begin{eqnarray}
{\dot Y}_k&=&+c\sqrt{k}\bigl[\cos(\omega t+\varphi)\cos\omega_0t~X_{k-1}-\cos(\omega t+\varphi)\sin\omega_0t~Y_{k-1}\bigr]\nonumber\\
& &+c\sqrt{k+1}\bigl[\cos(\omega t+\varphi)\cos\omega_0t~X_{k+1}+\cos(\omega t+\varphi)\sin\omega_0t~Y_{k+1}\bigr]\nonumber\\
& &+d\sqrt{k}\bigl[\sin(\omega t+\varphi)\cos\omega_0t~Y_{k-1}+\sin(\omega t+\varphi)\sin\omega_0t~X_{k-1}\bigr]\nonumber\\
& &-d\sqrt{k+1}\bigl[\sin(\omega t+\varphi)\cos\omega_0t~Y_{k+1}-\sin(\omega t+\varphi)\sin\omega_0t~X_{k+1}\bigr]\nonumber\\
& &+a_1\cos^2(\omega t+\varphi)~X_k+b_1\sin^2(\omega t+\varphi)~X_k,
\end{eqnarray}
\end{subequations}
where $a_1$, $b_1$, $c$ and $d$ have been defined as
\begin{subequations}
\begin{equation}
a_1=\frac{\alpha^2\omega_0^2}{2m\hbar(\omega_0^2-\omega^2)^2}, \quad\quad b_1=\frac{\alpha^2\omega2}{2m\hbar(\omega_0^2-\omega^2)^2}
\end{equation}
\begin{equation}
c=\frac{\alpha\omega_0^2}{\omega_0^2-\omega^2}\frac{1}{\sqrt{2m\hbar\omega_0}},\quad\quad d=\frac{\alpha\omega}{\omega_0^2-\omega^2}\sqrt{\frac{\omega_0}{2m\hbar}}.
\end{equation}
\end{subequations}
\newpage\noindent
For the resonant case (r), one can in addition make the following change of coefficients
\begin{equation}
D_k(t)=e^{-i\alpha^2 t^3/24m\hbar}{\widetilde D}_k(t)
\end{equation}
to eliminate the quadratic time dependence appearing in the expression (\ref{wr}) and (\ref{at}). Note that $D_k(0)={\widetilde D}_k(0)$ and $|D_k(t)|^2=|{\widetilde D}_k(t)|^2$. Doing the same
as it was done above, the real and imaginary parts of these new coefficients, ${\widetilde D}_k=\widetilde X_k+i\widetilde Y_k$, obey the equations
\begin{subequations}\label{eqr}
\begin{eqnarray}
\dot{\widetilde X}_k&=&+f(t)\widetilde Y_k-\sqrt{k}\biggl\{\bigl[h(t) \widetilde X_{k-1}-g(t) \widetilde Y_{k-1}\bigr]\sin\omega_0t+\bigl[h(t)\widetilde Y_{k-1}+g(t)\widetilde X_{k-1}\bigr]\cos\omega_0t\biggr\}\nonumber\\
& &+\sqrt{k+1}\biggl\{\bigl[h(t)\widetilde X_{k+1}+g(t)\widetilde Y_{k+1}\bigr]\sin\omega_0t-\bigl[h(t)\widetilde Y_{k+1}-g(t)\widetilde X_{k+1}\bigr]\cos\omega_0t\biggr\}
\end{eqnarray}
\begin{eqnarray}
\dot{\widetilde Y}_k&=&-f(t)\widetilde X_k+\sqrt{k}\biggl\{\bigl[h(t) \widetilde X_{k-1}-g(t) \widetilde Y_{k-1}\bigr]\cos\omega_0t-\bigl[h(t)\widetilde Y_{k-1}+g(t)\widetilde X_{k-1}\bigr]\sin\omega_0t\biggr\}\nonumber\\
& &+\sqrt{k+1}\biggl\{\bigl[h(t)\widetilde X_{k+1}+g(t)\widetilde Y_{k+1}\bigr]\cos\omega_0t+\bigl[h(t)\widetilde Y_{k+1}-g(t)\widetilde X_{k+1}\bigr]\sin\omega_0t\biggr\}
\end{eqnarray}
\end{subequations}
where the functions $f$, $h$, and $g$ have been defined as
\begin{subequations}
\begin{equation}
f(t)=\frac{\alpha^2}{8m\hbar\omega_0}\sin^2\omega_0t+\frac{\alpha^2t}{4m\omega_0\hbar}\cos(\omega_0t+\varphi)\sin(\omega_0t+\varphi),
\end{equation}
\begin{equation}
g(t)=\frac{\alpha}{2\hbar}\bigl[\frac{1}{\omega_0}\sin(\omega_0t+\varphi)+t\cos(\omega_0t+\varphi)\bigr],
\end{equation}
and
\begin{equation}
h(t)=\frac{\alpha\omega_0t}{2\hbar}\sin(\omega_0t+\varphi).
\end{equation}
\end{subequations}
The dynamical systems (\ref{eqnr}) and (\ref{eqr}) are solved by Runge-Kutta method a 4th-order. 
\newpage
\section{ Analytical Approach for $H(x,p,t)$}
The Hamiltonian of the forced harmonic oscillator is \cite{landau}
\begin{equation}
H(x,p,t)=H_0(x,p)+\alpha x\cos(\omega t+\varphi),
\end{equation}
where $H_0$ is given by
\begin{equation}
H_0(x,p)=\frac{p^2}{2m}+\frac{1}{2}m\omega_0^2 x^2.
\end{equation}
The solution of the eigenvalue problem
\begin{equation}
H_0\Phi=E\Phi 
\end{equation}
is well known \cite{messiah}, and its solution is the same as (\ref{eigenho}). Therefore, to solve the Shr\"odinger's equation (\ref{sh}), one proposes a solution of the form
\begin{equation}
|\Psi(t)\rangle=\sum_{n=0}^{\infty}e^{-iE_n^{(0)} t/\hbar}D_k(t)|n\rangle,
\end{equation}
which, after substituting in the Shr\"odinger's equation, using the eigenvalues and the orthogonality between any two states, and making some rearranging , the following dynamical systems 
is brought about  for the real and imaginary parts of the coefficients, $D_k(t)=x_k(t)+i y_k(t)$,
\begin{subequations}\label{eqh}
\begin{eqnarray}
\dot x_k&=&-\lambda\bigl[\sqrt{k}~x_{k-1}-\sqrt{k+1}~x_{k+1}\bigr]\cos(\omega t+\varphi)\sin\omega_0t\nonumber\\
& &-\lambda\bigl[\sqrt{k}~y_{k-1}+\sqrt{k+1}~y_{k+1}\bigr]\cos(\omega t+\varphi)\cos\omega_0t
\end{eqnarray}
\begin{eqnarray}
\dot y_k&=&-\lambda\bigl[\sqrt{k}~y_{k-1}-\sqrt{k+1}~y_{k+1}\bigr]\cos(\omega t+\varphi)\sin\omega_0t\nonumber\\
& &+\lambda\bigl[\sqrt{k}~x_{k-1}+\sqrt{k+1}~x_{k+1}\bigr]\cos(\omega t+\varphi)\cos\omega_0t,
\end{eqnarray}
\end{subequations}
where the constant $\lambda$ has been defined as
\begin{equation}
\lambda=\alpha\sqrt{\frac{\hbar}{2m\omega_0}}.
\end{equation}
These equations are also solved by using Runge-Kutta method at 4th-order.
\section{Boltzmann-Shannon entropy and energy}
Besides the probability to find the system in the state $|n\rangle$ at the time ``t'', $|D_k(t)|^2$, for the analysis of 
the dynamics of the system in the spaces ($x,\hat v$) and ($x,\hat p$), one can also consider the Boltzmann-Shannon entropy,
\begin{equation}\label{bosh}
S(t)=-\sum_{k=0}^l|D_k(t)|^2\ln|D_k(t)|^2,
\end{equation} 
and its average over a evolution time ``T'',
\begin{equation}\label{bsav}
\bar{S}=\frac{1}{T}\int_0^TS(t)dt,
\end{equation}
as parameter which characterize the quantum dynamics of the system. This parameter gives us an indication of how many states enter 
in the dynamics evolution of the system. Therefore, it gives an indication of the information lost in the dynamics due to the increasing of
the entropy in the quantum system.
\vskip1pc\noindent
In addition, one can also consider
the expectation value of the energy
\begin{equation}\label{ene}
\langle E\rangle(t)=\langle\Psi|\begin{pmatrix}\widehat H_0\\ \widehat K_0\end{pmatrix}|\Psi\rangle=\hbar\omega_0\sum_{n=0}^ln|D_n(t)|^2+\frac{1}{2}\hbar\omega_0,
\end{equation}
and its average value over the evolution time of the system,
\begin{equation}\label{ave}
\bar E=\frac{1}{T}\int_0^T\langle E\rangle(t)dt.
\end{equation}
This parameter gives information about how the energy is distributed among the states and how many of them are involved in the quantum dynamics.
\vskip1pc\noindent
In this way, solving the dynamical systems  (\ref{eqnr}), (\ref{eqr}, and (\ref{eqh}), the evolution of the probabilities $|D_k(t)|^2$'s are gotten. 
Thus, the Boltzmann-Shannon entropy (\ref{bosh}), the expectation value of the energy (\ref{ene}) and their average values (\ref{bsav}) and (\ref{ave}) 
can be calculates and can be compared for the quantization in the spaces ($x,v$) and ($x,p$). 
\section{Results}
One considers a proton with mass $m=1.6726219\times 10^{-27}Kg$ oscillating with a frequency $\omega_0=2\pi\times 10^9 Hz$ on a one dimensional line, and interacting with a periodic force force of amplitude
$\alpha=10^{-13}Newtons$ with frequency $\omega$ and phase $\varphi=0$. The initial conditions of the system are
\begin{equation}
C_k(0)=D_k(0)=\widetilde D_k(0)=\delta_{k0}, \quad k=0,\dots,11,
\end{equation}
that is, the system is on the ground state, and one selects ten exited possible state of the system.
\vskip1pc\noindent
For the non resonant case ($\omega\not=\omega_0$), the resulting dynamics from expressions (\ref{sh}) and (\ref{sk}) are exactly the same. There is not excitation of the system at all since
the system remains in the ground state in both cases.
\vskip1pc\noindent
For the resonant case ($\omega=\omega_0$), Figure 1 shows the probabilities of having the system on the ground state ($k=0$) and on the first excited state ($k=1$) for the quantization on the space ($x,v$), solid lines, and the quantization on the space ($x,p$), dotted lines. As one can see, for the Hamiltonian quantization approach (H) there are much more oscillations of the probabilities than the quantization of the constant of motion approach (K), that is, there are more transitions per unit time in the H-approach case than in the K-approach case. One must note that the probability to have the system in the first excited state for the K-approach case is totally different from the H-approach case. 
\begin{figure}[H]
\includegraphics[scale=0.5]{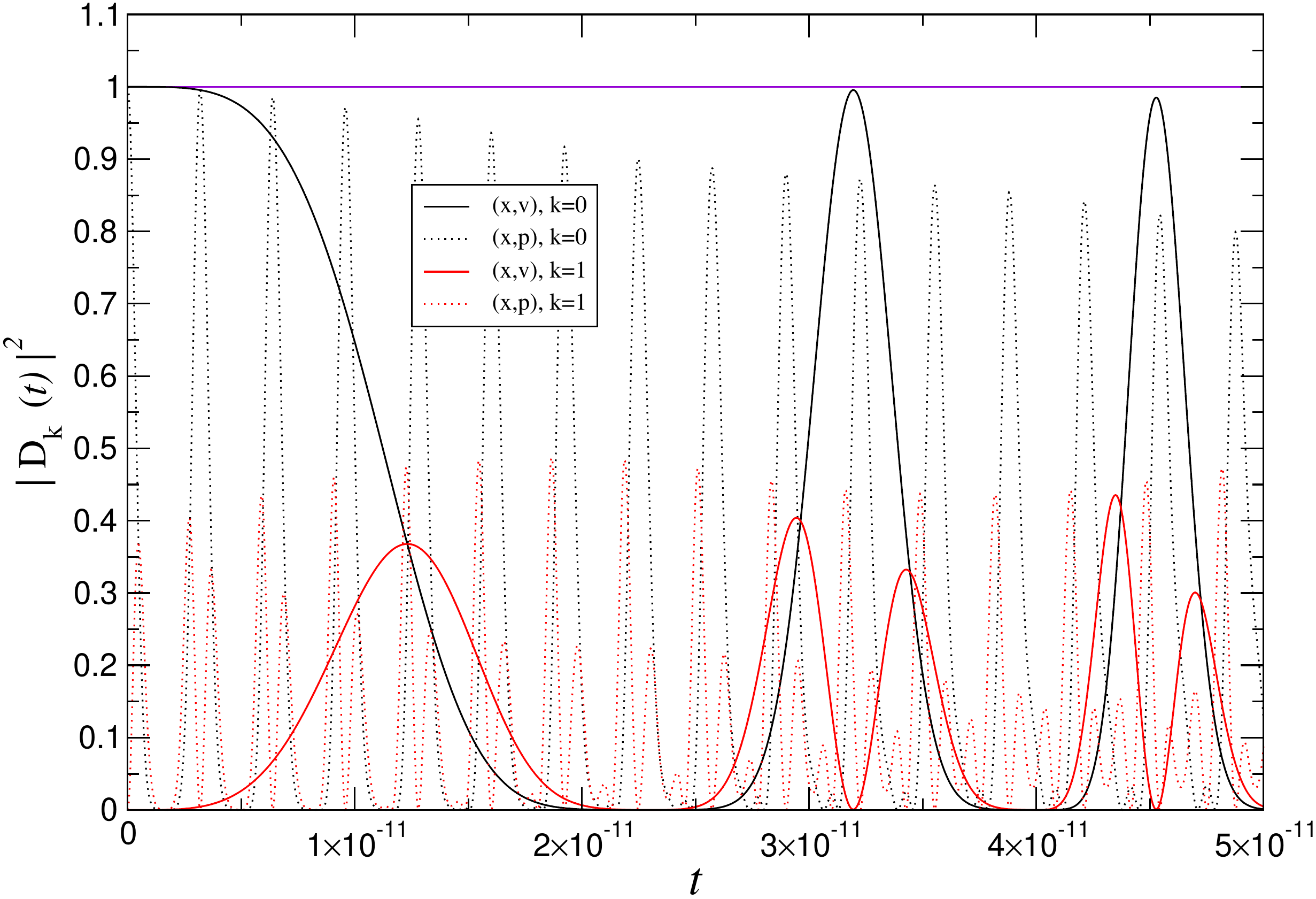}
\centering
    \caption{Ground state and first excited state evolution}
\end{figure}
Figure 2 shows the average value of the energy as a  function of the strength of the forced force ($\alpha$), and as one can see, this value is always higher for H-approach case than for the K-approach case. Although the difference is quite small and maybe out of experimental resolution. 
\begin{figure}[H]
\includegraphics[scale=0.5]{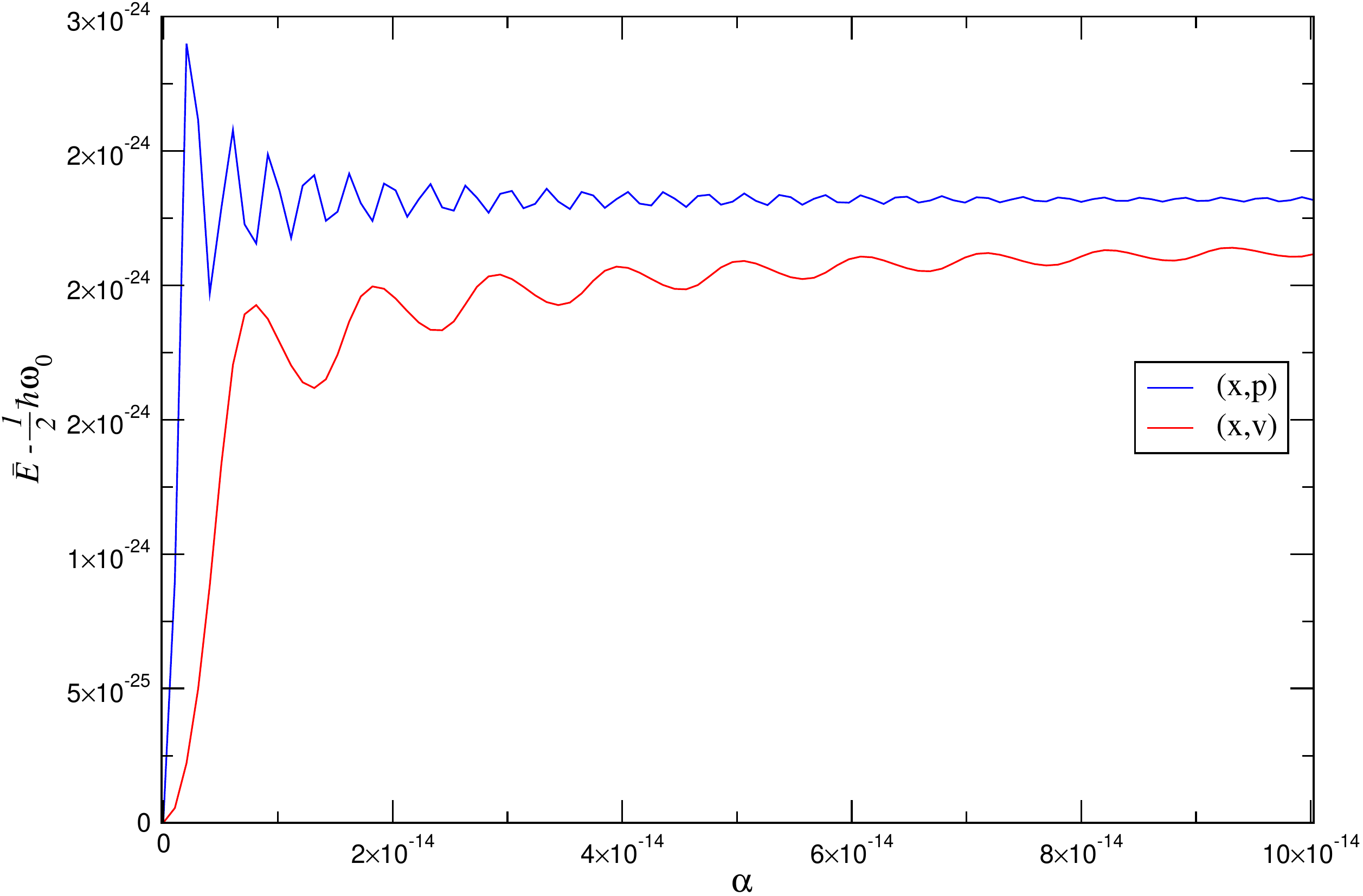}
\centering
    \caption{Average energy of the system}
\end{figure}
Figure 3 shows the average value of the Boltzmann-Shannon entropy as a function of the strength of the forced force ($\alpha$). Having total number of 11 states, the possible maximum entropy is 2.398. As the previous case. This parameter is always higher for the H-approach case than for the K-approach case. However, this difference is not so small and maybe could be used as a good parameter for experimental proposes. This difference means that the H-approach case  brings about more complex behavior in the quatum dynamics than the K-approach case, and that the H-approach case losses more information than the K-approach case. 
\begin{figure}[H]
\includegraphics[scale=0.5]{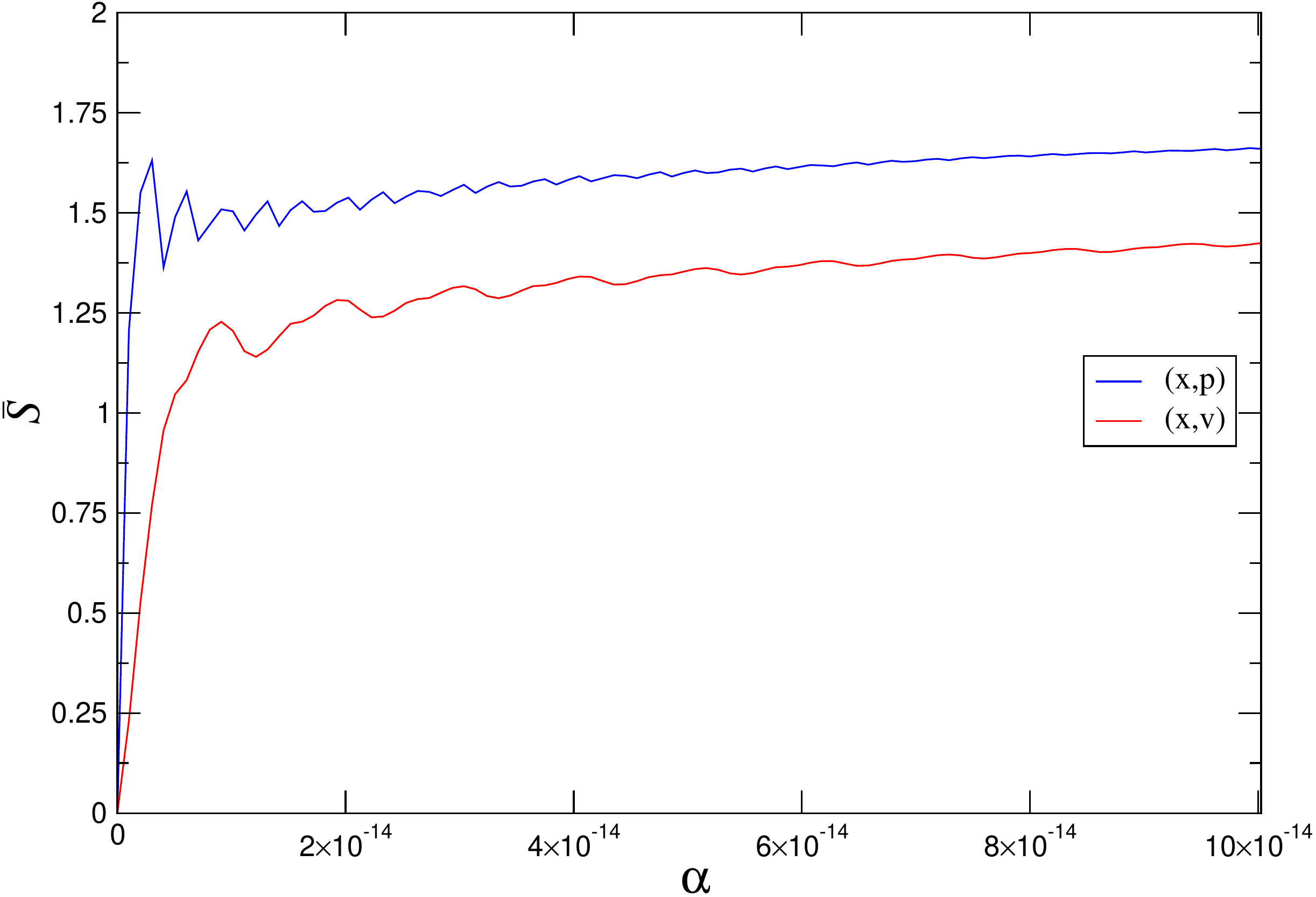}
\centering
    \caption{Avarage Boltzmann-Shannon entropy}
\end{figure}
\section{Conclusion}
The quantization of the 1-D forced harmonic oscillator was carried out with the operators ($x,\hat v$) using the assigned linear operator to a constant of motion $K(x,v,t)$ of the classical case. The restriction imposed on this constant was that it must reduced to the known energy expression when the forced force is zero. This quantization was compared with the usual quantization 
with the operators ($x,\hat p$) and the associated Hamiltonian $H(x,p,t)$ of the classical case. It was shown that the probabilities to find the system in the state $|n\rangle$, $|D_n(t)|^2$, has less oscillations in the K-quantization than in the H-quantization. In addition, the average values of the energy and the average value of the Boltzmann-Shannon entropy are lower in the K-quantization than in the H-quantization. Since the difference in the average value of the energy is quite small, this parameter does not look good to measure it experimentally. However, the difference in the entropy is significant and it represents a good parameter to look experimentally.   
\newpage
\bibliographystyle{unsrt} 
 \bibliography{bibliografia}

\begin{thebibliography}{10}

\bibitem{neumann}
J.~von Neumann.
\newblock {\em Mathematical Foundations of Quantum Mechanics}, volume 183.
\newblock Princeton University Press, 1955. Reprinted in paperback form., 1932.

\bibitem{born}
M.~Born.
\newblock Quantenmechanik der stossvorg\"ange.
\newblock {\em Z. Physik}, 38:803, 1926.

\bibitem{schr}
E.~Schr\"odinger.
\newblock An undulatory theory of the mechanics of atoms and molecules.
\newblock {\em Phys. Rev.}, 28:1049--1070, Dec 1926.

\bibitem{spielman}
I.~Spielman.
\newblock Quantum theory verified by experiment.
\newblock {\em Nature}, 545:293--294, 2017.

\bibitem{darboux}
G.~Darboux.
\newblock {\em Lecons sur la Th\'eorie G\'en\'erale des Surfaces,}, volume~3.
\newblock Gauthier-Villars, Paris, 1894.

\bibitem{goldstein}
H.~Goldstein.
\newblock {\em Classical Mechanics}, volume I, II.
\newblock Addison-Wesley, 1978.

\bibitem{lpez2007}
L\'opez~P. L\'opez~G.V. and L\'opez X.E.
\newblock Statistical physics on the space (x, v) for dissipative systems and
  study of an ensemble of harmonic oscillators in a weak linear dissipative
  medium.
\newblock {\em International Journal of Theoretical Physics}, 46(5):1100, Feb
  2007.

\bibitem{lll}
L\'opez~P. L\'opez~G.V. and L\'opez X.E.
\newblock Ambiguities on the hamiltonian formulation of the free falling
  particle with quadratic dissipation.
\newblock {\em Advanced Studies inTheoretical Physics}, 5(253), 2011.

\bibitem{dodman1}
Man'ko~V.I. Dodonov~V.V. and Skarzhinsky V.D.
\newblock The inverse problem of the variational claculas and the nonuniqueness
  of the quantization of classical systems.
\newblock {\em Hadronic Journal}, 4:1734, 1981.

\bibitem{dodman2}
Man'ko~V.I. Dodonov~V.V. and Skarzhinsky V.D.
\newblock Classically equivalent hamiltonians and ambiguities of quatization: A
  particle in a magnetic field.
\newblock {\em Il Nuovo Cimento B}, 69:185, 1982.

\bibitem{lop3}
Griselda~A. L\'opez~G.V. and Mart\'{\i}nez-Prieto R.M.
\newblock Ambiguity appearing on the hamiltonian formulation of quantum
  mechanics.
\newblock {\em Journal of Applied Mathematical and Physics}, 6:1382, 2018.

\bibitem{mont}
M.~Motensinos and G.F. Torres~del Castillo.
\newblock Symplectic quantization, inequivalent quatum theories, and
  heisenberg's principle of uncertainly.
\newblock {\em Physical Review A}, 70:032104--1, 2004.

\bibitem{lopez}
G.V. L\'opez.
\newblock Ambiguities appearing in the study of time-dependent constants of
  motion for the one-dimensional harmonic oscillator.
\newblock {\em Journal of Theoretical Physics}, 37:1617--1623, 1998.

\bibitem{lok}
Lopez G.V.
\newblock On the quantization of ome-dimensional conservative system of
  variable mass.
\newblock {\em Journal of Modern Physics}, 3:777--785, 2012.

\bibitem{lok2}
L\'opez G. and L\'opez P.
\newblock Velocity quantization approach of the one-dimensional dissipative
  harmonic oscillator.
\newblock {\em International Journal of Theoretical Physics}, 45:753, 2006.

\bibitem{dirac}
P.~Dirac.
\newblock A new notation for quantum mechanics.
\newblock 35(3):416--418, 1939.

\bibitem{burk}
Leventhal~J.J. Burkhardt~C.E.
\newblock {\em Harmonic Oscillator Solution Using Operator Methods.}
\newblock 2008.

\bibitem{landau}
E.~M.~Lifshitz Lev Davidovich~Landau.
\newblock {\em Course of Theoretical Physics}, volume~I.
\newblock 1970.

\bibitem{messiah}
A.~Messiah.
\newblock {\em Quatum Mechanics}, volume~I.
\newblock North Holland, John Wiley \& Sons. Ch. XII, 1966.

\end{thebibliography}

\end{document}